%% file: main.tex
\documentclass[runningheads]{llncs}
\usepackage[T1]{fontenc}
\usepackage{siunitx}
\usepackage{graphicx}
\usepackage{todonotes}
\usepackage{amsmath}
\usepackage{booktabs}
\usepackage{float} 
\usepackage{xcolor}
\usepackage{xspace}
\definecolor{forestGreen}{RGB}{46, 111, 64}
\definecolor{cherryRed}{RGB}{210,10,46}
\definecolor{glaciousBlue}{RGB}{103,141,198}

\newcommand\tool{\textsc{HammerWatch}\xspace}

\begin{document}
\title{Towards Remote Attestation of Microarchitectural Attacks: The Case of Rowhammer}
\titlerunning{Towards Remote Attestation of Microarchitectural Attacks} 

\authorrunning{Herrmann et al.}

\author{Martin Herrmann\inst{1}\orcidID{0009-0000-1953-8343} \and
Oussama Draissi\inst{1}\orcidID{0009-0005-6065-8087} \and
Christian Niesler\inst{1}\orcidID{0000-0002-8589-5231} \and
Ahmad-Reza Sadeghi\inst{2}\orcidID{0000-0001-6833-3598} \and
Lucas Davi\inst{1}\orcidID{0000-0002-5236-9067}}

\institute{paluno -- The Ruhr Institute for Software Technology,\\
University of Duisburg-Essen, Essen, Germany\\
\email{\{martin.herrmann, oussama.draissi, christian.niesler, lucas.davi\}@uni-due.de}
\and
Technical University of Darmstadt, Darmstadt, Germany\\
\email{ahmad.sadeghi@trust.tu-darmstadt.de}}

\maketitle               

\begin{abstract}
Microarchitectural vulnerabilities increasingly undermine the assumption that hardware can be treated as a reliable root of trust.
Prevention mechanisms often lag behind evolving attack techniques, leaving deployed systems unable to assume continued trustworthiness.
We propose a shift from prevention to detection through microarchitectural-aware remote attestation.
As a first instantiation of this idea, we present \tool, a Rowhammer-aware remote attestation protocol that enables an external verifier to assess whether a system exhibits hardware-induced disturbance behavior.
\tool leverages memory-level evidence available on commodity platforms, specifically Machine-Check Exceptions (MCEs) from ECC DRAM and counter-based indicators from Per-Row Activation Counting (PRAC), and protects these measurements against kernel-level adversaries using TPM-anchored hash chains.

We implement \tool on commodity hardware and evaluate it on \num{20000} simulated benign and malicious access patterns.
Our results show that the verifier reliably distinguishes Rowhammer-like behavior from benign operation under conservative heuristics, demonstrating that detection-oriented attestation is feasible and can complement incomplete prevention mechanisms.




\end{abstract}
\input{sections/introduction}

\input{sections/background}
\input{sections/problem-statement}
\input{sections/threat-model}

\input{sections/design.tex}

\input{sections/implementation.tex}

\input{sections/evaluation.tex}

\input{sections/related-work}

\input{sections/security_discussion.tex}
\input{sections/conclusion.tex}

\section*{Acknowledgments}
This work was partially funded by the Deutsche Forschungsgemeinschaft (DFG, German Research Foundation) through SFB 1119 (Project Number 236615297), Project S2, and by the DFG under the Priority Program SPP 2253 Nano Security (Project RAINCOAT, Project Number 440059533).

\input{sections/appendix.tex}

\newpage

%
%
%
\bibliographystyle{splncs04}
\bibliography{refs}

\end{document}

%% file: sections/introduction.tex
\section{Introduction}

System security rests on hardware as a root of trust.
Software-based protections, from operating system isolation to integrity measurement, assume that the underlying hardware reliably preserves memory integrity~\cite{akter2023survey}.
This assumption has guided decades of system design.
Yet modern performance optimizations increasingly introduce microarchitectural vulnerabilities that undermine this foundation~\cite{deng2022processor,lou2021survey,mishra2025modern}.
Features designed to improve latency, throughput, or density can unintentionally expose fault behavior or side effects that are exploitable below the software layer.

Prominent examples include cache timing channels~\cite{weissteiner2026continuous} and speculative execution attacks~\cite{kocher2020spectre,lipp2018meltdown}, which exploit access times and speculative execution to leak sensitive information.
Similarly, increasing DRAM density has introduced disturbance errors, called Rowhammer, that can be triggered through carefully crafted memory access patterns, to induce bit-flips enabling privilege escalation from unprivileged code to kernel-level execution~\cite{gruss2016rowhammer,kim2014flipping,kogler2022half}.
These attacks do not merely leak information; they can directly corrupt memory, violating isolation guarantees assumed by all higher software layers.
Because the corruption originates below the operating system, traditional integrity checks fail to detect or attribute the compromise and leave no reliable software-visible trace~\cite{gruss2018another}.


A decade of defensive research has shown that eliminating such vulnerabilities in deployed hardware is difficult.
Mitigations are often reactive, vendor-specific, and incomplete.
In the case of Rowhammer, countermeasures such as Target Row Refresh (TRR) and ECC memory have been overwhelmed by many-sided and Half-Double attacks~\cite{frigo2020trrespass,hassan2021uncovering,kogler2022half} or through multi-bit faults~\cite{cojocar2019exploiting,kwong2020rambleed}.
Even DDR5's Per-Row Activation Counting (PRAC) remains vulnerable~\cite{qureshi2025moat,woo2025qprac}.
Hardware-level fixes are costly and cannot protect already-deployed systems~\cite{gautam2019row,marazzi2023rega}.
Attack techniques continue to outpace defenses~\cite{baek2025marionette,luo2023rowpress,luo2024experimental,van2018guardion,woo2025dapper}, leaving systems unable to assume continued trustworthiness after deployment.

This persistent gap motivates a shift from preventing Rowhammer to detecting it.
Once a system is compromised, local detection becomes unreliable because an attacker with kernel-level privileges can suppress or forge any software-based monitoring~\cite{seaborn2015exploiting}.
Trust decisions must therefore be made externally.
Remote attestation provides this capability by enabling a verifier to assess system integrity through cryptographically protected evidence anchored in tamper-resistant hardware~\cite{coker2011principles}.
A Trusted Platform Module (TPM) can record measurements in a manner that remains unforgeable even to kernel-level adversaries~\cite{abera2016things}:

Yet existing remote attestation protocols cannot detect such attacks.
Software-based schemes such as IMA~\cite{sailer2004design} and Keylime~\cite{ruffin2025towards} measure boot-time and runtime software integrity.
Control-flow attestation protocols including C-FLAT~\cite{cflat_davi} and LO-FAT~\cite{dessouky2017fat} verify execution paths but assume correct underlying memory.
Hardware-anchored approaches using TPM~\cite{abera2016things} or TEEs~\cite{lee2020keystone} protect attestation evidence but still measure only software state.
Collectively, these mechanisms target firmware, boot loaders, and OS components, not hardware fault behavior.
To our knowledge, no attestation protocol systematically incorporates microarchitectural fault indicators into integrity evaluation.

In this paper, we explore how remote attestation can be extended to account for microarchitectural attacks.
We instantiate this concept using Rowhammer as a representative and well-studied hardware-level disturbance attack.
Rowhammer enables corruption of security-critical data structures such as page tables through DRAM disturbance effects~\cite{kim2014flipping,seaborn2015exploiting,kogler2022half}.
Its longevity, practical exploitability, and continued evolution make it an ideal case study for attesting to microarchitectural fault behavior.

Our prototype implementation, called \tool, constitutes the first Rowhammer-aware remote attestation protocol.
\tool integrates hard\-ware-level evidence, specifically Machine-Check Exceptions (MCEs) from ECC memory and Alert Back-Off (ABO) events from PRAC, into a TPM-anchored attestation architecture.
Our contributions are:
\begin{itemize}
    \item We design and implement a microarchitectural-aware remote attestation protocol, instantiated for Rowhammer, that captures memory-fault indicators from commodity hardware, protects measurements through TPM-anchored hash chains, and enables external verification of Rowhammer-related behavior.
    Our design ensures measurement integrity, authenticity, and freshness even against kernel-level privileged adversaries.
    Our implementation integrates MCE collection from the Linux EDAC subsystem and PRAC-based detection using Ramulator 2.0~\cite{luo2023ramulator}.
    \item We analyze the detection coverage of \tool against state-of-the-art Rowhammer variants, including Half-Double~\cite{kogler2022half}, many-sided attacks~\cite{frigo2020trrespass}, and ECCploit-style exploitation~\cite{cojocar2019exploiting}, as well as the related Rowpress vulnerability~\cite{luo2023rowpress}.
    \item We evaluate \tool on \num{20000} benign and malicious access patterns and demonstrate that the verifier reliably distinguishes Rowhammer-like behavior from benign operation under conservative detection heuristics.
\end{itemize}

%% file: sections/background.tex
\section{Background}

\paragraph{Rowhammer.}
DRAM stores data in capacitor-based cells organized into rows and columns, with cells requiring periodic refresh to retain their charge~\cite{kim2014flipping}.
Rowhammer exploits the physical proximity of DRAM rows: repeatedly activating an aggressor row causes charge leakage in adjacent victim rows, inducing bit-flips if the charge drops below the sensing threshold before refresh~\cite{kim2014flipping,seaborn2015exploiting}.
These bit-flips can corrupt security-critical data structures such as page tables, enabling privilege escalation~\cite{gruss2018another}.
Susceptibility has worsened with increasing DRAM density due to reduced capacitor sizes and tighter row spacing~\cite{kim2020revisiting,mutlu2019rowhammer}.

\paragraph{Error Correcting Code (ECC) and Machine-Check Exceptions (MCE).}
ECC memory detects and corrects bit errors by storing redundant parity information alongside data~\cite{fakhrzadehgan2022safeguard}.
Commodity ECC DRAM typically implements Single Error Correction, Double Error Detection, which can correct single-bit errors and detect two-bit errors per word~\cite{cojocar2019exploiting}.
When the memory controller detects an uncorrectable error or observes repeated correctable errors, it raises a MCE~\cite{cojocar2019exploiting}.
On Linux, MCEs are logged by the EDAC (Error Detection and Correction) kernel subsystem and include metadata such as error type and memory address.
While ECC can be bypassed through carefully templated multi-bit-flips~\cite{cojocar2019exploiting}, the templating process itself triggers detectable errors~\cite{di2023copy}.

\paragraph{Per-Row Activation Counting (PRAC).}
PRAC is a Rowhammer mitigation mechanism standardized in DDR5~\cite{JEDEC_JESD79-5C_2024}.
It maintains per-row activation counters within DRAM and triggers an Alert Back-Off (ABO) signal when a row exceeds a predefined activation threshold~\cite{qureshi2025moat,woo2025qprac}.
Upon receiving an ABO alert, the memory controller pauses normal operation and issues Refresh Management (RFM) commands to refresh potential victim rows~\cite{qureshi2025moat}.
Although PRAC-based prevention has been shown to remain vulnerable to advanced attacks~\cite{qureshi2025moat,woo2025qprac}, ABO events provide a valuable signal indicating excessive row activation, which we leverage for detection.

\paragraph{Remote Attestation (RA).}
RA enables an external verifier to assess system integrity through cryptographically protected evidence~\cite{coker2011principles}.
In a typical challenge-response protocol, the verifier sends a nonce and the prover responds with signed measurements of its internal state~\cite{ambrosin2020collective}.
Hardware-based RA anchors trust in a Trusted Platform Module (TPM), which protects cryptographic keys and records measurements in Platform Configuration Registers (PCRs)~\cite{abera2016things}.
PCRs are extended by cumulatively hashing new measurements into the previous value, creating an append-only chain that prevents forgery or rollback even by privileged attackers~\cite{arthur2015platform}.
Existing RA mechanisms attest software state such as firmware and OS components, but do not capture hardware fault behavior.

%% file: sections/problem-statement.tex
\section{Problem Statement}

Rowhammer defenses span over refresh-based mitigation such as TRR, access-pattern monitoring, throttling, memory displacement, ECC, memory isolation, and DRAM architectural changes, yet attackers have repeatedly adapted to and bypassed these mechanisms~\cite{baek2025marionette,cojocar2019exploiting,frigo2020trrespass,hassan2021uncovering,kogler2022half}.
Prevention-focused approaches often incur substantial performance, memory, or design overhead~\cite{fakhrzadehgan2022safeguard}.
Many rely on disturbance-locality assumptions that do not hold universally or remain vulnerable to advanced strategies such as many-sided hammering~\cite{frigo2020trrespass}, Half-Double~\cite{kogler2022half}, ECC-aware flip patterns~\cite{cojocar2019exploiting}, and coupled-row effects~\cite{baek2025marionette}.
Furthermore, some of these techniques are difficult to to deploy universally~\cite{gautam2019row,marazzi2023rega}.

As a consequence, even systems equipped with state-of-the-art mitigations cannot safely assume that Rowhammer has been fully neutralized at the hardware or firmware level, especially on newer DRAM generations and heterogeneous deployment settings even on the latest DDR5~\cite{meyer2025phoenix}.

This situation exposes a central research gap: existing work predominantly treats Rowhammer as a phenomenon to be suppressed locally, rather than as a security signal that can inform post-deployment trust decisions. Current defenses rarely integrate Rowhammer-induced behavior into system-wide integrity assessments and therefore offer limited means to detect, attribute, and contain successfully exploited machines in the field. In particular, no prior work has systematically explored how Rowhammer-specific effects could be modeled and incorporated into remote attestation protocols, so that external verifiers can reason about Rowhammer-induced compromise and react accordingly. Recent defenses such as SafeGuard~\cite{fakhrzadehgan2022safeguard} are usually integrity protection mechanisms. This work addresses that gap by investigating Rowhammer-aware remote attestation as a complementary protection layer: instead of attempting to eliminate Rowhammer entirely, we aim to expose and encode Rowhammer-relevant behavior within attestation, enabling external detection, classification, and containment of compromised systems even in the presence of imperfect or partially bypassed in-situ mitigations.

%% file: sections/threat-model.tex
\section{Threat Model}
\label{sec:threat-model}

We consider a distributed infrastructure composed of multiple interconnected systems that exchange sensitive data or participate in security-critical operations. In such environments, individual systems must be trusted before they are allowed to interact with other components of the infrastructure.
\tool assumes that this trust decision is enforced externally and that remote attestation is used as a prerequisite for interaction.

We assume that an attacker can compromise individual systems using Row\-hammer attacks.
In particular, the attacker may gain unprivileged DRAM access and induce bit-flips to escalate privileges and obtain kernel-level control.
We explicitly do not aim to prevent such compromises.
Once a Rowhammer attack succeeds, software-level security mechanisms on the affected system can no longer be trusted, and the system is considered compromised.
However, we assume the presence of a TPM that cannot be compromised and can securely protect cryptographic keys and attestation state. Furthermore, we assume that the attacker has no physical access to the compromised system.

The goal of \tool is therefore not to protect a single machine, but to limit the impact of compromised systems within a larger infrastructure.
By enabling remote parties to detect evidence of Rowhammer activity, \tool allows compromised systems to be identified and isolated.
This prevents them from interacting with other systems or accessing sensitive data.
Only systems that successfully complete attestation are permitted to participate in security-critical operations.
This reduces the risk of lateral movement, data exfiltration, and prolonged undetected abuse.



%% file: sections/design.tex
\section{Design Overview}
\label{design}

This section introduces the overall design of \tool, our Rowhammer-aware remote attestation framework.
We define Rowhammer-specific measurement sources and then describe how these measurements are protected, transmitted, and evaluated within a principled attestation workflow.

Our design follows a general principle: microarchitectural attacks often manifest through hardware-level anomaly signals that can be externally verified, such a elevated branch mispredictions for speculative execution attacks or cache miss rates for cache timing attacks.
We therefore structure our framework around capturing, protecting, and attesting such hardware-derived measurements. We instantiate this design using Rowhammer as a concrete and well-studied disturbance attack.

\paragraph{Measurements for Rowhammer Detection.}
Remote attestation (RA) relies on system measurements that allow a verifier to evaluate integrity. In traditional RA, these measurements capture system configuration and software state; the same concept applies in this context.


We identify two measurement sources for detecting Rowhammer activity. Counter-based measurements monitor whether row activations exceed predefined thresholds, indicating potential bit-flip-inducing behavior. ECC-based measurements capture mismatches between stored and recomputed codes, revealing bit-flips after they occur. Prior work demonstrates that combining both methods enhances detection coverage, as they provide complementary protection~\cite{cojocar2019exploiting,luo2023rowpress}.


Other approaches, such as probabilistic TRR, memory isolation, or purely hardware-level mitigations, do not expose measurable signals suitable for attestation. Because these mechanisms offer no observable evidence of active attacks, we exclude them. 


Determining whether a Rowhammer attack has successfully compromised a system remains challenging. Bit-flips may result from benign effects or occur during templating rather than exploitation, and excessive memory accesses may not always imply success. Therefore, we conservatively classify any detected Rowhammer activity that produces bit-flips as indicative of compromise.

\paragraph{Measurement Protection - Integrity and Authenticity}
Attestation measurements must be protected against manipulation by an attacker with kernel-level privileges.
Storing measurements in regular memory is insufficient, as they can be modified, deleted, or fabricated.
Even storing measurements in a secure module is problematic if integrity is not cryptographically enforced using a practically unforgeable append only hash.

We therefore rely on a trusted hardware module that continuously extends a cumulative hash over all measurements.
Each new measurement is hashed into the previous hash, forming an append-only chain following the Merkle--Damg{\aa}rd construction.
Similar hash chains were used by previous research in a different context, such as control-flow attestation~\cite{cflat_davi}.

Both the hash extension and storage are performed exclusively inside the trusted module.
The module signs the current cumulative hash using a private key stored internally.
The verifier stores the corresponding public key.

As a result, any valid attestation response necessarily includes the measurement history up to the point of compromise.
While an attacker may suppress future measurements, earlier Rowhammer activity cannot be removed from the hash chain.
This property exposes compromised systems during attestation, even after an attacker has attained kernel-level privileges.

\paragraph{Availability and Confidentiality}

The absence of a correct attestation response is treated as a strong indicator of compromise or malfunction.
Reduced availability only limits the attacker’s ability to interact with other systems and therefore does not require additional countermeasures, as it is in the attackers self-interest to not undermine availability to the verifier.

Measurement confidentiality is not strictly required for proper attestation, but it impedes attackers from reverse engineering defense parameters.
Measurements stored on the prover are therefore encrypted.
Encryption protects measurements even if the attacker gains kernel-level access prior to attestation.

\paragraph{TOCTTOU Attack Prevention.}
Time-of-check-to-time-of-use (TOCTTOU) attacks pose a significant threat to remote attestation.
An attacker could collect valid attestation evidence before launching a Rowhammer attack.
This stale evidence could later be used to deceive the verifier in a replay attack.
To ensure freshness, the verifier includes a nonce in the attestation request.
The trusted module signs the nonce together with the cumulative measurement hash.
This binds the evidence to a specific attestation instance and prevents replay.

\begin{figure}[tb]
    \centering
    \includegraphics[width=1\textwidth]{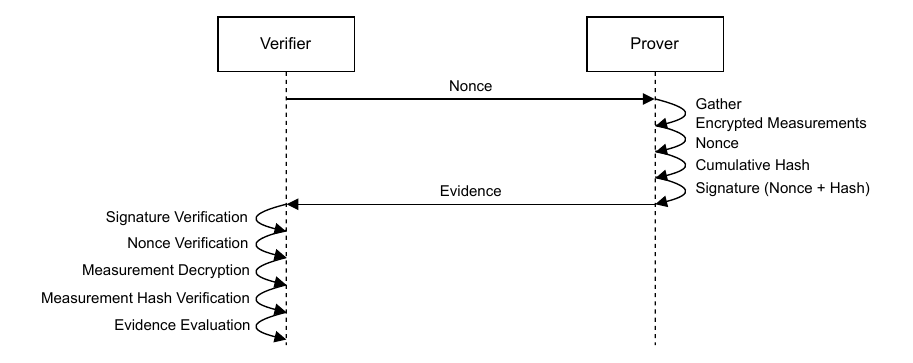}
    \caption[Conceptual RA challenge-response protocol]{\textbf{Conceptual RA challenge-response protocol:} The verifier issues a nonce with the attestation request. The prover responds with encrypted measurements, the nonce, and the corresponding signature. The verifier checks these elements to determine whether the prover is compromised.}
    \label{fig:remote_attestation_protocol}
\end{figure}

\paragraph{Remote Attestation Protocol.} Our design follows a challenge–response protocol, as depicted in Fig.~\ref{fig:remote_attestation_protocol}.
The verifier initiates attestation by sending a nonce to the prover.
The prover responds with encrypted measurements, the nonce, the cumulative hash, and a signature over the nonce and hash.
The verifier verifies the signature, checks the nonce, decrypts the measurements, verifies the integrity of the measurements, and evaluates whether the prover is compromised.

\paragraph{Evidence Evaluation}
Deciding whether the evidence provided by the prover is sufficient to prove its integrity is non-trivial.
Previous research has not established a clear consensus on which heuristics are most effective.
In counter-based detection, the proposed thresholds vary widely~\cite{canpolat2024breakhammer,qureshi2022hydra}. 
Recent work focuses on the DDR5 DRAM standard \cite{qureshi2025moat,woo2025qprac}, which specifies per-row counters but leaves threshold selection to vendors, who do not disclose the values used by their products \cite{JEDEC_JESD79-5C_2024}.
Moreover, the heuristics and defenses, that we are aware of, are not always able to reliably distinguish between benign and malicious row accesses.

Similarly, distinguishing between benign and malicious bit-flips in ECC-based detection is challenging.
While accidental flips can occur under commodity workloads, such events are rare and highly dependent on specific conditions \cite{loughlin2022moesi}.
Based on this insight, Di Dio et al. \cite{di2023copy} consider three or more bit-flips in a row as an indication of a Rowhammer attack.

While heuristics for interpreting measurements remain an open research topic, certain protocol-level indicators provide clear evidence of compromise or malfunction.
A missing, severely delayed, syntactically incorrect, or incomplete attestation response strongly suggests compromise.
Likewise, an unexpected nonce or a mismatch between the signature and the nonce or hash are strong indicators of malicious tampering.

%% file: sections/implementation.tex
\section{Implementation}
\label{implementation}

We implemented a prototype of the proposed Rowhammer-aware remote attestation (RA) design using Ramulator~2.0~\cite{luo2023ramulator} and demonstrate feasibility for commodity platforms.
The implementation integrates mechanisms for collecting memory-level measurements, securing them against a powerful adversary, and providing them to a remote verifier through an attestation protocol.

Our prototype consists of three main components:
\begin{enumerate}
    \item Collection of ECC- and counter-based measurements on the prover,
    \item protection of these measurements with respect to integrity and confidentiality, and
    \item an end-to-end implementation of the RA challenge–response protocol.
\end{enumerate}

\paragraph{Measurement Types.}
As argued in Section~\ref{design}, we instantiate our framework using ECC- and counter-based detection mechanisms that expose Rowhammer-related disturbance behavior.
While research continues to improve these detection methods, most recent work on counter-based mechanisms has focused on prevention rather than detection \cite{qureshi2025moat,woo2025qprac}.
In particular, the Rowhammer prevention method called Per-Row Activation Counting (PRAC) was recently standardized by the Joint Electron Device Engineering Council (JEDEC) for DRAM.
Similarly, recent ECC research has primarily emphasized enhancing prevention rather than detection \cite{di2023copy}.

In the form of DDR5 DRAM with PRAC and ECC, these commodity systems lack the ability to prevent state-of-the-art Rowhammer exploits \cite{meyer2025phoenix}, but they still provide reasonably mature detection signals \cite{di2023copy,JEDEC_JESD79-5C_2024}.

\paragraph{Machine-Check Exceptions.}
Commodity ECC DRAM typically employs Single Error Correction, Double Error Detection (SECDED) \cite{di2023copy}.
Although SECDED can be bypassed through templating-based Rowhammer attacks~\cite{cojocar2019exploiting}, the templating process itself causes detectable bit-flips \cite{di2023copy}.
Detected memory errors are reported by the memory controller (MC) to the operating system via Machine-Check Exceptions (MCEs)~\cite{cojocar2019exploiting}.
On Linux, the reporting and handling of MCE is managed by the EDAC kernel module \cite{di2023copy}.
MCEs may also be generated for non-Rowhammer hardware faults, such as overheating, bus errors, or CPU cache errors.
While vendor dependent, MCEs typically include sufficient contextual information to serve as meaningful evidence in an RA protocol, such as type of error and memory address.
In our implementation, we collect MCEs from the systemd journal.

\paragraph{Per-Row Activation Counting (PRAC).}
PRAC builds on ideas introduced by Panopticon~\cite{bennett2021panopticon} and has been standardized in DDR5~\cite{JEDEC_JESD79-5C_2024}.
The underlying principles of PRAC remain consistent, although vendors have implemented some aspects differently, e.g., the threshold for counter-based refreshes.

PRAC implements per-row activation counters within DRAM \cite{qureshi2025moat}.
When excessive activations, surpassing the activation threshold, are detected, it triggers the Alert Back-Off (ABO) alert.
The underlying ABO protocol operates in the following four states \cite{nazaraliyev2025practical}:

\begin{enumerate}
\item \textbf{Normal} The DRAM operates normally.
\item \textbf{Pre-Recovery} When a row exceeds the activation threshold, the DRAM issues an ABO alert to the MC. Depending on the vendor, the MC has a window of up to 180 ns to perform activations before being blocked from issuing further activations.
\item \textbf{Recovery} The MC issues between one and four Refresh Management (RFM), where each RFM operation refreshes the rows with the highest counters in each bank to protect potential victim rows.
\item \textbf{Delay} To prevent repeated memory locking, the next ABO alert can only be issued after a predefined number of activations, after which the protocol returns to the \textbf{Normal} state.
\end{enumerate}

PRAC and ABO operate entirely within the DRAM and memory controller without software exposure.
Since implementing a custom memory controller is out of scope, we follow prior work~\cite{qureshi2025moat,woo2025qprac} and simulate PRAC and ABO behavior.
For this purpose, we use Ramulator~2.0~\cite{luo2023ramulator}, which supports PRAC simulation and which is not available in other DRAM simulators such as Gem5~\cite{binkert2011gem5}.

\subsection{Securing Measurements}
Measurements are stored locally in a log file without access restrictions, as an attacker with kernel-level privileges can bypass conventional access controls.
Instead of relying on storage protection, we secure measurements cryptographically with respect to integrity and confidentiality.

\paragraph{Integrity.}
To ensure integrity under our threat model, we require a trusted hardware module supporting protected hash extension and enforcing signing policy.
In particular, the signing key must only be usable if the designated protected hash is included in the signed input, preventing attackers from signing fabricated measurement logs.
We use a Trusted Platform Module (TPM), which acts as a hardware root-of-trust, providing secure key storage and Platform Configuration Registers (PCRs).
TPMs use Non-volatile RAM (NVRAM), which is slower and more expensive than DRAM, but is considered immune to the Rowhammer bug and therefore regarded as secure against Rowhammer attacks \cite{gude2023defending}.

We dedicate a PCR, called PCR$[X]$, within the TPM for the RA protocol.
We concatenate the previous value PCR$[X]_{\text{old}}$ with each new measurement $m$, hash it using the hash function $H$, and store the result as a new value PCR$[X]_{\text{new}}$:

\begin{equation}
    \mathrm{PCR}[X]_{\text{new}} \leftarrow H(\mathrm{PCR}[X]_{\text{old}} \| m)
    \label{PCReq}
\end{equation}

The hash algorithm used is SHA-256.
This construction yields an append-only record of all prior measurements and is computationally infeasible to forge~\cite{arthur2015platform}.

An attacker who gains kernel-level privileges may block future measurements, but cannot remove earlier entries from the PCR hash.
Since Rowhammer activity is required to obtain such privileges, the attack itself is already reflected in the PCR chain, rendering the attack futile, since its own activities are already captured in the PCR chain.
Replay attacks are prevented through nonce-based freshness guarantees.

TPMs have known limitations under strong adversaries, as kernel-level attackers can bypass policy enforcement and request signatures over arbitrary data.
These limitations allow the attacker to fabricate benign-looking measurements, hash them together in a different PCR or outside the TPM, and then instruct the TPM to sign the forged hash with its private key.
This would produce valid-looking evidence that falsely suggests the prover’s system is uncompromised.
To address this, we extend PCR$[X]$ with a secret value before logging any measurements.
This secret is issued by the verifier after each reboot and is deleted immediately after being hashed into the PCR.
Because the attacker never learns this secret, fabricating a valid PCR value becomes infeasible.

PCRs are reset on reboot~\cite{francillon2014minimalist}, meaning that the integrity of measurements obtained before the restart can no longer be guaranteed.
In our threat model, an attacker could persist on the prover’s system and exploit this reset to erase the log.
However, Rowhammer itself does not inherently aid an attacker in establishing persistence.
Therefore, and based on research on malware persistence detection \cite{aslan2020comprehensive}, we assume that persistence detection mechanisms are present on the prover’s system.
The reboot also requires the prover to re-request a fresh secret after reboot.
The verifier only issues a new secret if a reboot is detected, preventing attackers from simply requesting a new secret at any time to create a new fabricated and benign looking measurement log with matching PCR.

Finally, during attestation, the TPM signs the PCR value together with a nonce using RSA, ensuring authenticity and freshness.

\paragraph{Confidentiality.}
While confidentiality is not strictly required for proper attestation, exposing measurements could enable attackers to adapt future Rowhammer strategies.
Symmetric encryption was deemed unsuitable, as kernel-level attackers could misuse TPM-resident keys to decrypt measurements.
We therefore use asymmetric RSA encryption, allowing the prover to encrypt measurements with the verifier’s public key~\cite{afolabi2016comparative}.
Only the verifier can decrypt them, even if the prover is compromised.
Although RSA is slower than symmetric encryption~\cite{yassein2017comprehensive}, the number of measurements is small and performance is not critical. \title{do we need a cite here?}

\subsection{Remote Attestation Protocol Implementation}
\label{implementation:Remote_Attestation_Protocol_Implementation}

We implemented the RA protocol as a challenge–response protocol over TLS-secured TCP connections.

\begin{figure}[b]
    \centering
    \includegraphics[width=0.6\textwidth]{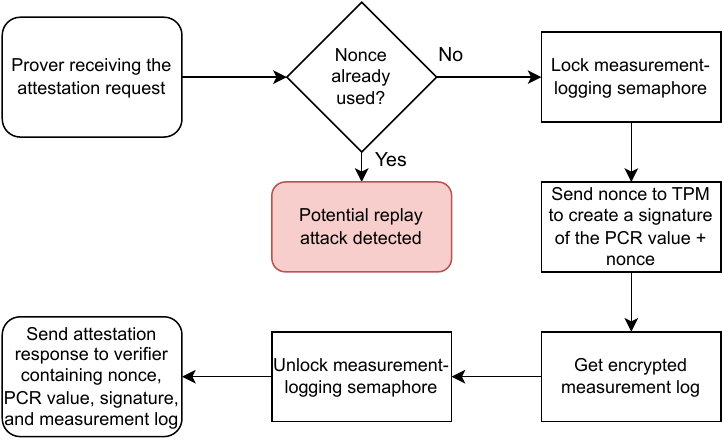}
\caption[Attestation Request Processing]{\textbf{Attestation Request Processing:} Flow describing how the prover processes an attestation request in the implemented attestation protocol.}
    \label{fig:prover_request_processing}
\end{figure}

Upon receiving the request, see Fig.~\ref{fig:prover_request_processing}, the prover verifies the nonce to prevent replay attacks. 
After confirming the nonce hasn’t been used yet, measurement logging is halted using a semaphore in order to avoid race conditions between extending the PCR value and appending the measurement log, and using them as evidence.
Without this safeguard, a measurement could occur during evidence calculation, which may extend the PCR value without being appended to the measurement log.
Thereby creating a mismatch between the log and the PCR value in the evidence, potentially leading to the false conclusion that the prover is compromised.
Next, the TPM signs the PCR together with the nonce.
The prover then collects the evidence consisting of the nonce, the PCR value, the signature, and the encrypted measurement log.
Afterwards the semaphore is released and the attestation response is sent to the verifier.

\begin{figure}[tb]
    \centering
    \includegraphics[width=1\textwidth]{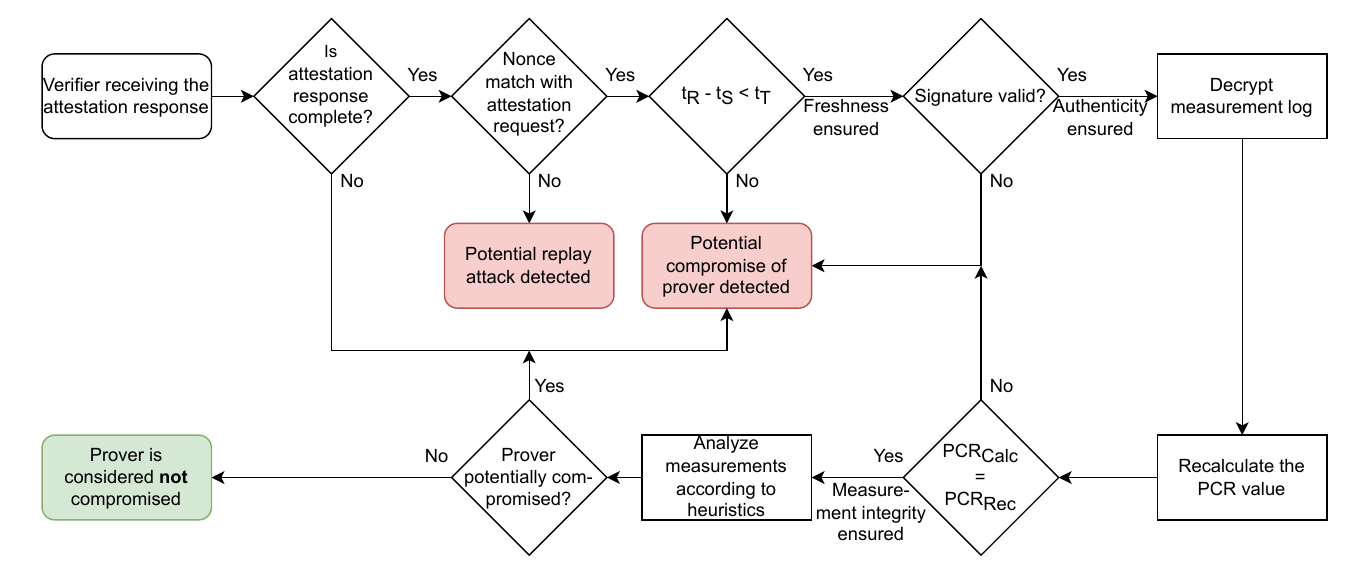}
    \caption[Attestation Response Processing]{\textbf{Attestation Response Processing:} Detailed flow of how the verifier processes an attestation response in the implemented attestation protocol. 
    Here, $t_R$ denotes the time of receiving the attestation response, $t_S$ the time of sending the attestation request, and $t_T$ the timeout threshold.
    $PCR_{Calc}$ refers to the recalculated PCR value, and $PCR_{Rec}$ refers to the PCR value received in the attestation response.}
    \label{fig:verifier_response_processing}
\end{figure}

Once the verifier receives the complete attestation response, see Fig.~\ref{fig:verifier_response_processing}, it compares the nonce with the one used in the original request and checks whether the time between sending the attestation request and receiving the attestation response exceeds the timeout threshold.
A nonce mismatch indicates a potential replay attack, while an attestation response exceeding the timeout threshold indicates abnormal protocol execution, which under our threat model may result from a compromised prover deliberately delaying the response after receiving the challenge to conduct a TOCTTOU attack.
In our implementation, we used a simple timeout threshold of 10 seconds; however, this value should be refined depending on the network conditions and the computing power of the prover.
If both the nonce and response time are valid, the freshness of the attestation response is ensured.

Next, the verifier checks the signature using the corresponding public key to ensure the authenticity of the prover’s TPM.
The measurement log is then decrypted with the associated private key.
The initial secret PCR value is extended with the measurements, one by one, starting from the reset value zero.
The resulting hash, i.e., the recalculated PCR value, must match the PCR value provided in the evidence; if so, the measurements are considered intact.

Based on the verified measurements and any irregularities, such as delayed responses, signature mismatches, nonce mismatches, or PCR mismatches, the verifier determines the state of the prover.
Due to the lack of consensus on Rowhammer detection heuristics, we adopt a conservative policy:
The prover is considered uncompromised only if the measurements contain fewer than three MCEs and fewer than three ABO alerts.
This follows the suggestion by Di Dio et al. \cite{di2023copy}, see Section~\ref{design}.
However, this heuristic can be easily adapted if required.
Finally, the TLS/TCP connection is closed, thereby concluding the RA.
Based on the assessed state of the prover, a decision whether to continue interacting with the prover or not is made.

\subsection{Implementation Summary}

\begin{figure}[tb]
    \centering
    \includegraphics[width=1\textwidth]{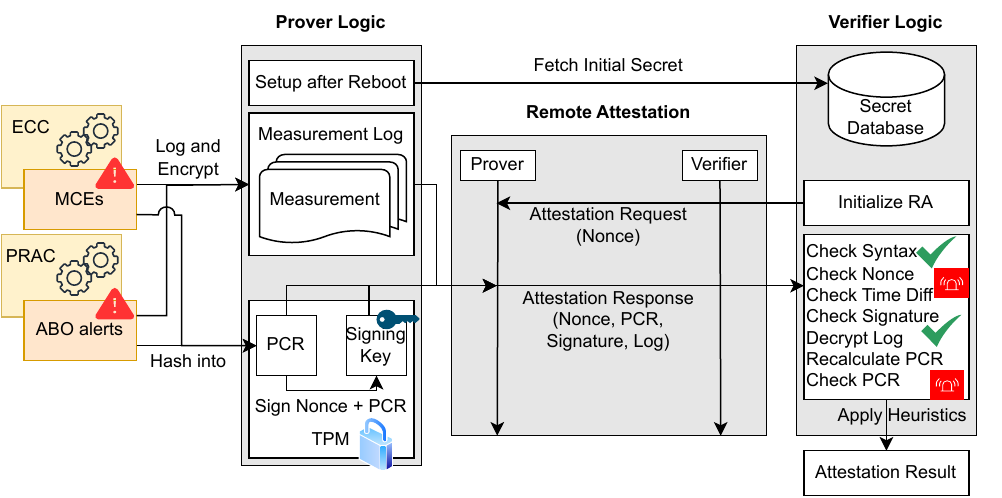}
    \caption[Overview of the HammerWatch architecture.]{\textbf{Overview of the HammerWatch architecture.} 
    }
    \label{fig:overall_implementation}
\end{figure}

The overall implementation architecture is depicted in Fig.~\ref{fig:overall_implementation}.
The prototype consists of 2,008 lines of Python and shell scripts.
It relies on standard Linux tools (e.g., openssl, base64), Python libraries (socket, threading, secrets) and tpm2-tools from Intel\footnote{\url{https://github.com/tpm2-software/tpm2-tools}}.
Ramulator~2.0~\cite{luo2023ramulator} and mce-inject\footnote{\url{https://github.com/andikleen/mce-inject}} were used for testing.
We additionally implemented a test-pattern generator to produce benign and hammering access traces compatible with Ramulator’s YAML format.

%% file: sections/evaluation.tex
\section{Evaluation}

\tool represents the first prototype instantiation of our mic\-ro\-arch\-i\-tec\-tu\-ral-aware attestation framework, using Rowhammer as a representative fault attack.
We evaluate our prototype Rowhammer-aware remote attestation (RA) protocol with respect to detection accuracy, overhead, and scalability.
We further analyze additional attack surfaces introduced by our design.

\paragraph{Rowhammer Detection Accuracy.}
We evaluated detection accuracy by simulating 20,000 benign and malicious row-access patterns on the prover.
For each run, measurements were collected and the RA protocol was executed, after which the verifier classified the prover as either compromised or uncompromised.

Counter-based measurements were obtained by simulating PRAC and ABO behavior in Ramulator~2.0~\cite{luo2023ramulator}, which supports PRAC modeling.
Ramulator~2.0 processes lists of row-access commands and reports DRAM behavior, including triggered ABO alerts.
Since publicly available Rowhammer traces compatible with Ramulator are limited, we implemented a dedicated generator to produce PRAC-compatible access patterns representing both benign and hammering behavior.
ECC-related measurements were simulated using mce-inject~\cite{andikleen_mceinject_2024}.

Following Section~\ref{implementation:Remote_Attestation_Protocol_Implementation}, we classify runs as malicious if more than three ABO alerts or more than three MCEs are observed~\cite{di2023copy}.

We conducted 10,000 benign and 10,000 malicious runs, with PRAC-enabled DDR5 DRAM modeled in Ramulator and consistent with the DDR5 standard~\cite{JEDEC_JESD79-5C_2024}.
The full configuration is provided in Appendix~\ref{appendix_ramulator2_config}.
To avoid rebooting the system between runs, we extended the RA implementation with an evaluation mode that resets the prover log while preserving PCR verification semantics.

Across all runs, the verifier correctly classified every benign run as uncompromised and every malicious run as compromised, resulting in no false positives or false negatives.
These results demonstrate reliable detection of Rowhammer-related behavior in the evaluated configuration.

\paragraph{Limitations.}
However, the validity of these results depends on the heuristics and measurements employed.
Since robust Rowhammer detection criteria remain an open research problem, future work should focus on refining heuristics and incorporating additional measurements.
Our design explicitly supports such extensions without requiring architectural changes.

Due to limited availability of PRAC-compatible datasets and lack of hardware exposure of PRAC signals, we rely on self-generated access patterns with a full-system simulation, consistent with prior work~\cite{di2023copy,qureshi2025moat,woo2025qprac}.


\paragraph{Overhead and Scalability.}
Because PRAC and MCEs are simulated and logging is workload-independent, precise runtime overhead quantification is infeasible.
Similarly, RA incurs only limited overhead, especially since RA is usually performed infrequently and is workload-independent.
As a result, precise overhead quantification is infeasible.

Nevertheless, qualitative estimates are possible.
On the prover, overhead stems from logging, PCR extension, measurement encryption, and participation in the RA protocol.
The CPU impact is minimal, and memory is dominated by the measurement log, while cryptographic keys require less than 10 KB.
%
%
The verifier incurs negligible overhead outside attestation and performs computation only during verification, and secret distribution adds minimal additional cost.


Scalability is primarily determined by measurement volume.
Under typical workloads, Rowhammer-indicating events are rare~\cite{loughlin2022moesi}, resulting in low overhead.
%
%
In data center environments, where Rowhammer-like behavior and memory faults are more common~\cite{meza2015revisiting}, scalability may be more challenging and warrants further study.

Key management across many provers can be addressed by deploying a centralized verifier.
Communication overhead from transmitting large logs is manageable with TLS/TCP, though excessively large logs could be abused for denial-of-service attacks, which we discuss below.


\paragraph{Additional Attack Surfaces.}
\tool introduces potential TOCTTOU and Denial-of-Service (DoS) attack surfaces.
Nonce-based freshness largely prevents TOCTTOU attacks.
Although an attacker could theoretically compromise the system between request and response, this window is small, below 0.5~seconds in our experiments, and insufficient for a full Rowhammer attack.
Moreover, as with all RA schemes, integrity is only guaranteed at attestation time.

DoS attacks could target both prover and verifier by flooding them with attestation requests.
TLS authentication and nonce validation mitigate this threat.
A compromised prover could attempt to overwhelm the verifier with excessively large measurement logs.
Preventing this is non-trivial. 
A defense mechanism is to treat oversized logs as evidence of compromise, though this risks false positives for workloads that generate many benign measurements.
Alternatively, limiting the log to include only measurements that are individually strong indicators of compromise would allow the verifier to stop processing early, once sufficient evidence is gathered.
However, this would significantly restrict the types of measurements usable for \tool and only slightly reduce the decryption computational cost.



%% file: sections/related-work.tex
  \section{Related Work}
  \label{related work}
  
\paragraph{Rowhammer Defenses.}
  Early Rowhammer defenses focused on reducing the attacker’s available hammering window by increasing DRAM refresh rates~\cite{kim2014flipping,seaborn2015exploiting}.
  While this approach can reduce the probability of bit-flips, it is ineffective against optimized attacks and incurs significant performance overhead~\cite{bhati2015dram}.
  Similarly, attempts to disable cache flush instructions to hinder cache eviction can be bypassed through alternative access patterns~\cite{aweke2016anvil}.
  More advanced defenses focus on detecting hammering behavior based on characteristic access patterns.
  Probabilistic refresh mechanisms refresh neighboring rows with a certain probability upon activation~\cite{kim2014flipping}. 
  Counter-based defenses monitor row activations within a refresh interval and classify rows as aggressors once predefined thresholds are exceeded~\cite{lee2019twice}.
  However, selecting appropriate thresholds becomes increasingly difficult with newer DRAM generations and advanced attack techniques requiring fewer activations~\cite{luo2023rowpress,mutlu2019rowhammer}.
  Numerous optimizations aim to reduce counter overhead, including coarse-grained counters~\cite{taouil2021lightroad}, dynamic hot-row grouping~\cite{seyedzadeh2018mitigating}, approximate counting algorithms~\cite{park2020graphene}, Bloom-filter-based tracking~\cite{bostanci2024comet}, and hardware-based designs such as Panopticon~\cite{bennett2021panopticon}.
  Despite near-ideal granularity, even hardware-assisted counter-based approaches remain vulnerable to optimized attacks and incur non-trivial design complexity~\cite{meyer2025phoenix,qureshi2025moat,woo2025qprac}.
  
  Once hammering is detected, most defenses attempt to prevent bit-flips through Target Row Refresh (TRR), throttling, or memory displacement.
  Target Row Refresh (TRR) refreshes rows adjacent to suspected aggressors to restore the victim rows’ charge but have been shown bypassable by advanced variants~\cite{hassan2021uncovering,qureshi2022hydra}.
  Memory throttling, instead, limits either the number of activations to the aggressor row or the overall memory accesses of the aggressor process to prevent an attacker from conducting sufficient hammering operations to induce bit-flips within a refresh interval~\cite{canpolat2024breakhammer}.
  Memory displacement restricts an attacker’s ability to access victim-adjacent rows by remapping the aggressor row to a different random row~\cite{saileshwar2022randomized} or isolated quarantine area~\cite{saxena2022aqua} to remove direct adjacency to victim rows.
  Error Correcting Codes (ECC), such as SECDED~\cite{peterson1972error} and Chipkill~\cite{dell1997white}, increase resilience but can be bypassed or overwhelmed by carefully crafted flip patterns~\cite{cojocar2019exploiting}.
  Stronger integrity schemes incur significant storage and performance overhead~\cite{fakhrzadehgan2022safeguard}.
  
\paragraph{Memory Isolation.} techniques attempt to physically separate attacker-con\-trolled and victim data.
These include guard rows, i.e., empty rows placed between isolated memory regions~\cite{tatar2018throwhammer,van2018guardion}, privilege-based isolation~\cite{brasser2017can}, and VM-level separation~\cite{loughlin2023siloz}. These approaches incur moderate to significant memory overhead and rely on assumptions about disturbance locality that do not universally hold~\cite{baek2025marionette}.
However, isolation mechanisms have been bypassed through Half-Double~\cite{kogler2022half}, implicit hammering~\cite{zhang2020pthammer}, and speculative execution attacks~\cite{zhang2020ghostknight}.
Hardware-based defenses modify DRAM architectures to address Rowhammer at its root cause.
Examples include per-row hardware counters~\cite{bennett2021panopticon,qureshi2025moat,woo2025qprac}, additional sense amplifiers to allow for parallel row refreshes~\cite{marazzi2023rega}, and physical mitigation techniques such as the introduction of nanowires to reduce electron migration~\cite{gautam2019row}. While effective in principle, these approaches are expensive, impractical to retrofit, and do not protect already deployed systems.  

\paragraph{Rowhammer Attack Advancements.}
As defenses improved, attackers developed techniques specifically designed to evade them.
Double-sided, many-sided, and non-uniform hammering reduce the number of activations per row while staying below detection thresholds~\cite{frigo2020trrespass}.
Half-Double exploits TRR itself by inducing secondary aggressors, which allows an aggressor to be one row further away~\cite{kogler2022half}.
ECC-based defenses can be bypassed by inducing exploitable flip patterns or overwhelming correction capabilities in so-called ECCploit-like attacks~\cite{cojocar2019exploiting}.
Memory isolation has been circumvented using speculative execution~\cite{zhang2020ghostknight}, implicit hammering~\cite{zhang2020pthammer}, Half-Double~\cite{kogler2022half}, and coupled-row behavior~\cite{baek2025marionette}.
Coupled-row attacks violate the assumption that Rowhammer affects only adjacent rows and enable attacks across subarrays~\cite{baek2025marionette}.
Rowhammer has been demonstrated across architectures, platforms, and threat models.
Attacks exist on x86 and ARM~\cite{kim2014flipping}, from JavaScript~\cite{gruss2016rowhammer}, on mobile devices~\cite{frigo2018grand}, via DMA~\cite{van2018guardion}, and remotely via RDMA~\cite{tatar2018throwhammer}.
Rowhammer has also been used for privilege escalation~\cite{gruss2018another}, data exfiltration~\cite{kwong2020rambleed}, denial of service~\cite{woo2025dapper}, cryptographic key extraction~\cite{poddebniak2018attacking}, and machine-learning weight extraction and model corruption~\cite{rakin2022deepsteal,yao2020deephammer}.


\paragraph{Rowpress.}
Rowpress represents a more recent class of fault attacks that induce bit-flips by holding DRAM rows open for extended periods rather than repeatedly activating them~\cite{luo2023rowpress}.
Rowpress relies on physical mechanisms distinct from Rowhammer, inducing electron migration from storage nodes toward the bitline through prolonged exposure to sustained electric fields~\cite{zhou2024unveiling}.
Unlike Rowhammer, Rowpress becomes more effective at higher temperatures and flips bits in the opposite direction~\cite{luo2023rowpress}.
The combination of Rowhammer and Rowpress can significantly increase the number of induced bit-flips~\cite{luo2024experimental}.
The capabilities of Rowpress allow attackers to reduce the required number of activations by up to two orders of magnitude, in extreme cases achieving bit-flips with a single activation.
As a result, Rowpress can bypass many counter-based, probabilistic, and even some hardware-assisted Rowhammer defenses.
Proposed defenses include limiting maximum row open times, lowering counter thresholds and combining them with ECC, and hardware mechanisms that disconnect rows from the row buffer after activation~\cite{luo2023rowpress,seongil2014row}.


%% file: sections/security_discussion.tex
\section{Security Considerations}

While our evaluation demonstrates that \tool can reliably detect Rowhammer attacks, the rapid evolution of attack techniques requires an assessment of its robustness against state-of-the-art variants.
In particular, advanced attacks such as Half-Double, many-sided Rowhammer, ECCploit-like attacks, coupled-row attacks, and Rowhammer-like fault mechanisms challenge traditional prevention-based defenses. 

Half-Double Rowhammer attacks hammer rows further away from the victim to evade preventative mechanisms and can even exploit TRR to amplify hammering effects~\cite{kogler2022half}.
Since our heuristics are largely location-independent and treat fault-related measurements as being compromised regardless of their origin, such attacks would still be detectable.
However, detection ultimately depends on the chosen heuristics. For instance, heuristics that only treat ABO alerts on rows directly adjacent to rows triggering MCEs as potentially compromised will struggle against Half-Double.

Many-sided Rowhammer attacks attempt to bypass counter-based defenses by distributing hammering activity across multiple rows~\cite{frigo2020trrespass}.
While this reduces per-row activation counts, these attacks do not avoid inducing bit-flips and therefore still trigger MCEs.
In contrast, ECCploit-like attacks rely on intensive hammering during templating to bypass ECC protections~\cite{cojocar2019exploiting}, which leads to ABO alerts due to the high row activation intensity.
By incorporating both PRAC-related ABO alerts and ECC-triggered MCEs, \tool is able to detect both attack classes.

Coupled-row attacks violate the assumption that disturbance effects are limited to immediately adjacent rows~\cite{baek2025marionette}.
These attacks still induce bit-flips and thus trigger MCEs.
Their ability to evade ABO alerts depends on vendor PRAC implementations, as counters maintained per physical wordline would still observe elevated activation behavior~\cite{baek2025marionette}.
Consequently, detection remains possible through MCEs and, depending on the implementation, also through ABO alerts.

Rowpress is not a Rowhammer attack but similarly induces bit-flips through sustained row activation~\cite{luo2023rowpress}.
Due to its low activation count, Rowpress rarely triggers ABO alerts.
However, the resulting memory faults still manifest as MCEs, allowing partial detection by \tool.

Despite this coverage, combinations of advanced attacks may pose challenges.
For example, using many-sided Rowhammer during the templating phase of ECCploit-like attacks could reduce both per-row activation counts and the number of triggered ABO alerts, depending on the PRAC threshold.
Although ECCploit-like attacks still trigger MCEs during templating, which would be detected, the obtained information could potentially enable future attacks that induce fewer observable faults.
While such combined attacks have not yet been demonstrated, they represent a realistic threat to both \tool and existing Rowhammer defenses and warrant further investigation.

\section{Discussion}

\paragraph{Future Work.}
Several directions exist for future work.
First, robust detection heuristics remain an open challenge.
There is no consensus on which indicators best characterize Rowhammer attacks, especially in the presence of evolving attack techniques.
Future work should systematically develop and validate heuristics that remain effective against emerging Rowhammer variants.

Second, practical evaluation is limited by current platform constraints.
PRAC-related information is not exposed outside the memory controller, and suitable datasets of real-world row access patterns are scarce.
Future work should therefore evaluate the protocol on physical hardware exposing Rowhammer-relevant signals and under realistic workloads. 

Third, attention should be paid to data center environments.
Such workloads differ from consumer workloads and are expected to generate a higher number of benign row accesses and memory faults.
This may require data center workload-specific heuristics and a reassessment of overhead and scalability.


\paragraph{Applicability Beyond Rowhammer.}
Although instantiated for Rowhammer, the framework is attack-agnostic.
Any microarchitectural vulnerability that produces measurable hardware side effects, such as abnormal performance counter activity, hardware error reports, or timing anomalies, can be integrated as attestation inputs on platforms exposing these signals.
The TPM-anchored logging mechanism and verifier-side heuristic evaluation remain unchanged.
This modularity enables extension to other microarchitectural attacks, such as speculative execution, cache timing attacks, and emerging microarchitectural attack classes, positioning microarchitectural-aware attestation as a general post-deployment trust assessment strategy.


%% file: sections/conclusion.tex
\section{Conclusion}

Microarchitectural vulnerabilities challenge the assumption that deployed hardware remains trustworthy.
Using Rowhammer as a representative disturbance-based attack, we shift the focus from prevention to detection and design a Rowhammer-aware remote attestation (RA) protocol that enables external verification of hardware-induced fault behavior to contain compromised systems.



Building on established RA principles, we integrate ECC-triggered Machine-Check Exceptions (MCEs) and PRAC-based Alert Back-Off (ABO) alerts into a TPM-anchored measurement architecture.
Our prototype demonstrates feasibility on commodity platforms, and the evaluation shows reliable detection under conservative heuristics with limited operational overhead.



This work demonstrates that remote attestation can be extended beyond software integrity to account for microarchitectural fault behavior.
Detection-oriented attestation offers a complementary strategy to incomplete prevention mechanisms and provides a foundation for post-deployment trust assessment in the presence of evolving hardware threats.



%% file: sections/appendix.tex

\clearpage
\appendix

\section{Ramulator 2.0 Configuration Parameters}
\label{appendix_ramulator2_config}
This section summarizes the Ramulator 2.0 configuration used in our simulations.
The simulator was customized to model a DDR5 memory system with PRAC-based refresh control.

\begin{table}[H]
\centering
\small
\begin{tabular}{ll}
\toprule
\textbf{Component} & \textbf{Parameter / Value} \\
\midrule
\multicolumn{2}{l}{\textit{Frontend Configuration}} \\
 & \texttt{impl = BHO3} \\
 & \texttt{clock\_ratio = 8} \\
 & \texttt{num\_expected\_insts = 1,000,000} \\
 & \texttt{llc\_capacity\_per\_core = 2 MB} \\
 & \texttt{llc\_num\_mshr\_per\_core = 16} \\
 & \texttt{inst\_window\_depth = 128} \\

\multicolumn{2}{l}{\textit{Address Translation}} \\
 & \texttt{impl = RandomTranslation} \\
 & \texttt{max\_addr = 16 GiB (17179869184 bytes)} \\

\multicolumn{2}{l}{\textit{Memory System}} \\
 & \texttt{impl = BHDRAMSystem} \\
 & \texttt{clock\_ratio = 3} \\

\multicolumn{2}{l}{\textit{Address Mapping Policy}} \\
 & \texttt{impl = RoBaRaCoCh\_with\_rit} \\
\bottomrule
\end{tabular}
\caption{System-Level Simulation Configuration}
\end{table}

The frontend models an out-of-order core with bounded instruction window and an LLC per core, while randomized address translation and the RoBaRaCoCh (Row-Bank-Rank-Column-Channel) mapping policy define physical row adjacency behavior relevant for disturbance-based attacks.




\begin{table}[H]
\centering
\begin{tabular}{ll}
\toprule
\textbf{Parameter} & \textbf{Value} \\
\midrule
\texttt{impl} & DDR5-VRR \\
\texttt{org.preset} & DDR5\_16Gb\_x8 \\
\texttt{org.channel} & 1 \\
\texttt{org.rank} & 2 \\
\texttt{timing.preset} & DDR5\_3200AN \\
\texttt{RFM.BRC} & 2 \\
\texttt{PRAC} & true \\
\bottomrule
\end{tabular}
\caption{DRAM Configuration}
\end{table}

The DRAM model instantiates a DDR5-3200 device with two ranks and PRAC support enabled.

\begin{table}[H]
\centering
\begin{tabular}{ll}
\toprule
\textbf{Parameter} & \textbf{Value} \\
\midrule
\texttt{impl} & PRACDRAMController \\
\texttt{BHScheduler.impl} & PRACScheduler \\
\texttt{RefreshManager.impl} & AllBank \\
\texttt{RowPolicy.impl} & ClosedRowPolicy \\
\texttt{RowPolicy.cap} & 4 \\
\bottomrule
\end{tabular}
\caption{DRAM Controller Parameters}
\end{table}

The controller uses a PRAC-aware scheduler with a closed-row policy and all-bank refresh strategy. The row cap limits repeated activations before forced closure.

\begin{table}[H]
\centering
\begin{tabular}{ll}
\toprule
\textbf{Parameter} & \textbf{Value} \\
\midrule
\texttt{abo\_delay\_acts} & 4 \\
\texttt{abo\_recovery\_refs} & 4 \\
\texttt{abo\_act\_ns} & 180 \\
\texttt{abo\_threshold} & 16 \\
\texttt{debug} & true \\
\bottomrule
\end{tabular}
\caption{PRAC Plugin Parameters}
\end{table}

These parameters configure the Alert Back-Off (ABO) mechanism, including activation delay, recovery behavior, and trigger thresholds. Debug mode was enabled to expose PRAC-related events for measurement collection.
